\documentclass[prb,a4paper,twocolumn,floatfix,showpacs,showkeys,amsmath,amssymb,nobibnotes,altaffilletter]{revtex4}

\usepackage{graphicx}% Include figure files
\usepackage{pslatex}
\usepackage{xspace}
\usepackage{subfig}

\begin{document}

\preprint{}

\title{Comment on ``Interface state recombination in organic solar cells''}

\author{Carsten Deibel}\email{deibel@physik.uni-wuerzburg.de}%
\author{Alexander Wagenpfahl}%
\affiliation{Experimental Physics VI, Julius-Maximilians-University of W{\"u}rzburg, 97074 W{\"u}rzburg, Germany}

\date{\today}

\begin{abstract}
In a recent paper, Street et al. [Phys. Rev. B 81, 205307 (2010)] propose first order recombination due to interface states to be the dominant loss mechanism in organic bulk heterojunction solar cells, based on steady-state current--voltage characteristics. By applying macroscopic simulations, we found that under typical solar cell conditions, monomolecular or bimolecular recombination cannot be inferred from the slope of the light intensity dependent photocurrent. In addition, we discuss the validity of calculating a mobility--lifetime product from steady-state measurements. We conclude that the experimental technique applied by Street et al.\ is not sufficient to unambiguously determine the loss mechanism. 
\end{abstract}

\pacs{73.50.Gr, 71.20.Rv}

\keywords{organic semiconductors; polymers; recombination}

\maketitle

Before we consider the paper by Street et al.\cite{street2010} by addressing the dependence of the photocurrent on the light intensity and the determination of the mobility--lifetime-product of photogenerated charges from steady-state current--voltage characteristics, we will briefly define terms needed for discussing recombination. We conclude our comment by a brief summary.

\section{Definitions}

The naming conventions for recombination mechanisms are not without ambiguity. As the order of decay and the number of involved charges are not necessarily identical, we briefly name the differences.

Considering the decay of charges by using the continuity equation for a particle (here hole) of density $p$, $dp/dt = G-R$ and its generation rate $G$ and recombination rate $R$. A generalised form of the latter is $R=k_\alpha p^\alpha$, where $k_\alpha$ is a prefactor and $\alpha$ the \emph{order of recombination}.

Monomolecular recombination involves only one particle, whereas bimolecular recombination involves two, for instance an electron and a hole. Auger recombination requires three particles. Sometimes, only \emph{free} charges are considered,\cite{chikao2004book} but for the sake of a more general treatment of the recombination type, we refer to free and trapped charges alike, and distinguish between them as described further below. Monomolecular recombination is a first order process, as the recombination rate is directly proportional to the density of that particle. Bimolecular recombination depends on the product of the density of both recombination partners. If electron density $n$ and hole density $p$ are within the same order of magnitude, $n\approx p$, the recombination rate $k_2 np\approx k_2 n^2$ describes a second order process. However, bimolecular recombination does not always have this appearance. For instance, if $n\gg p$, which may happen if a trapping mechanism for electrons exists, then this process can be a first order process: even if all holes have recombined, the remaining density of electrons has only been reduced by a fraction. Under these conditions, $R\approx p/\tau$ with $\tau=1/k_2'n$, which is approx.\ constant for $n\gg p$. We believe the term first-order bimolecular recombination to be appropriate, but due to the recombination rate only depending on one particle, Street et al.\cite{street2010} and others\cite{chikao2004book} refer to it as monomolecular recombination. Depending on the extent of trapping, also an order of recombination between one and two is possible. 

For bimolecular recombination, also recombination orders of higher than two have been observed,\cite{shuttle2008,juska2008,deibel2009} which has been assigned to the influence of trapping without the trapped charges being actively involved in the recombination process.\cite{zaban2003,foertig2009}---for instance if they are within a nanocrystal of the acceptor phase, and thus cannot be reached by the mobile oppositely charged particle. These traps are not necessarily due to impurities, but are a typical property of hopping systems with energetic and spatial disorder. These findings are compatible with the multiple-trapping-and-release model.\cite{arkhipov1982} To our knowledge, Auger recombination was never reported in disordered organic semiconductor systems.

The definition of geminate vs.\ nongeminate recombination is already given by Street et al.: the recombination of two particles from the same precursor state---such as an electron and a hole resulting from the dissociation of one exciton---is called geminate (and is a monomolecular first order process). The recombination of two particles from different precursor states---such as already free charges meeting for recombination---is a nongeminate process.

\section{Light Intensity Dependence of the Photocurrent}\label{sec:jsc}

\begin{table}[h]
	\centering
		\begin{tabular}{lll}
			\hline
			\hline
			parameter & value & description \\
			\hline
			$E_{g}$	& $1.05~\text{eV}$ & effective band gap\cite{vandewal2008, veldman2009} \\
			$\Phi_n$, $\Phi_p$ & $0.05$ eV & injection barriers\\
			$\mu_n,~\mu_p$ & $10^{-8}~\text{m}^2\text{/Vs}$ & mobilities\cite{baumann2008}\\
			$L$ & $100~\text{nm}$ & active layer thickness\\
			$T$ & $300~\text{K}$ & temperature\\
			$N_{c},~N_v$ & $ 10^{26}~\text{m}^{-3} $ & effective density of states\\		
			$\epsilon_r$ & $3.4$ & relative static permittivity\cite{persson2005}\\	
			\hline
			\hline
		\end{tabular}
	\caption{Parameters used in the macroscopic simulation.}
	\label{tab:param}
\end{table}

Applying a macroscopic device simulation,\cite{wagenpfahl2010,wagenpfahl2010a} we calculated the light--intensity dependent current--voltage characteristics by varying the generation of free charges over four orders of magnitude. The simulation parameters can be found in Table.~\ref{tab:param}. Our aim was to see how easily different polaron recombination mechanisms could be determined by evaluating the dependence of the photocurrent, in analogy to the inset of Fig.~10 of Street et al.\cite{street2010} There, a linear dependence is reported, i.e., a direct proportionality between photocurrent and the illumination intensity.

Assuming first order recombination, $R\approx n/\tau$, with a carrier lifetime $\tau$, we found that the scaling of the photocurrent vs.\ light intensity is linear, i.e., slope 1 behaviour (not shown), irrespective of the value of $\tau$.

\begin{figure}
	\includegraphics[width=9.0cm]{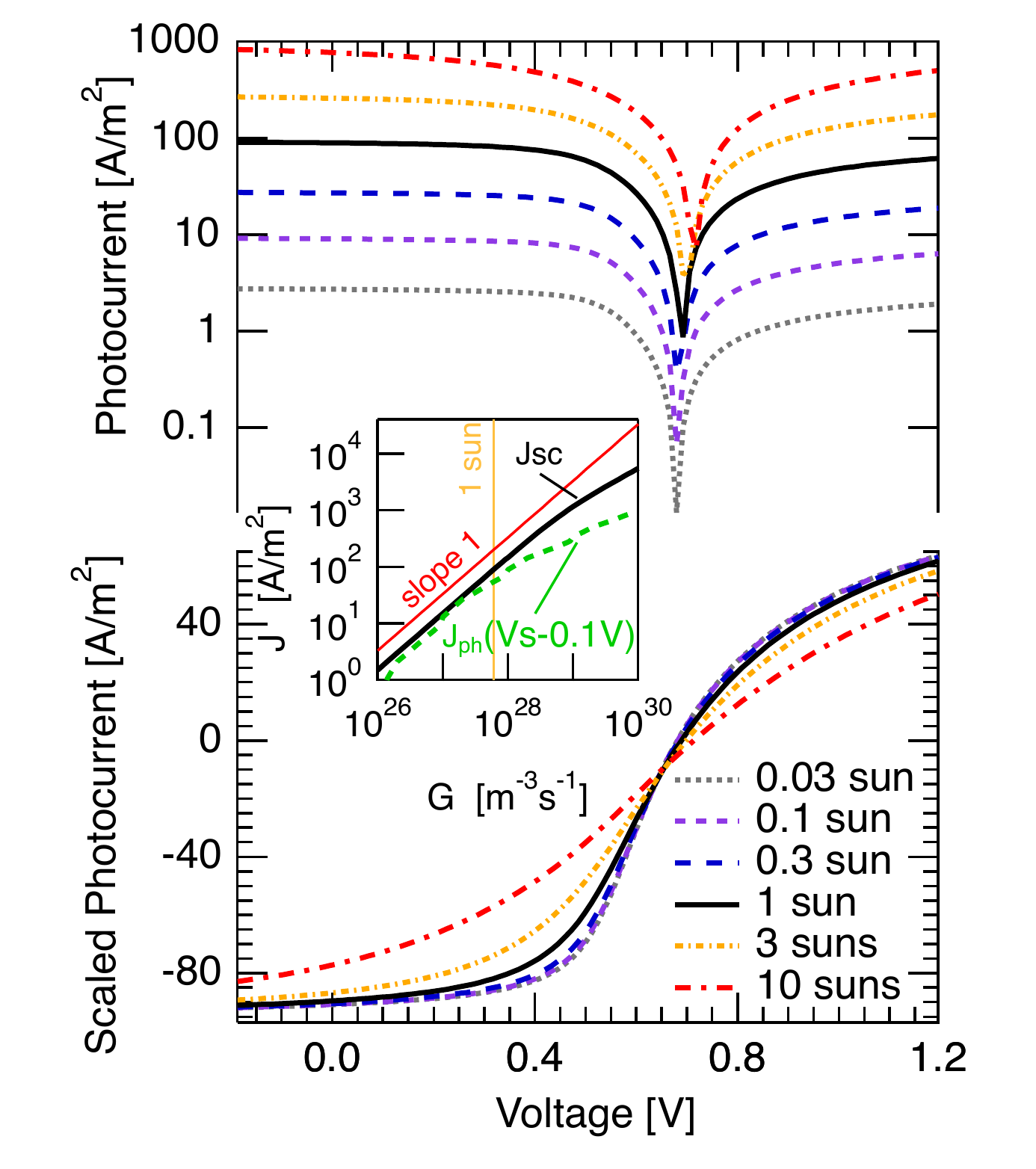}%
	\caption{(Color Online) Simulation of the photocurrent vs.\ voltage with bimolecular recombination ($\zeta=0.1$) for different illumination densities (0.03 suns to 10 suns). (top) The absolute photocurrent on a logarithmic scale.  (bottom) The photocurrent scaled to the generation rate. Here, it becomes clear that deviations from the shape of the characteristics are only seen above 1 sun, despite the recombination being bimolecular. (inset) The short circuit current and the photocurrent close to the quasi flatband voltage, $J_{ph}(V_s-0.1V)$. An illumination of 1 sun corresponds to $G=6\cdot 10^{27}$~m$^{-3}$s$^{-1}$, as indicated by the vertical line. The thin line indicates the slope 1 behaviour. Both current scale linearly with the generation rate up to at least one sun.\label{fig:j-rescaled}}
\end{figure}

This behaviour indeed changes if we assumed bimolecular (Langevin type) recombination, applying the recombination rate
\begin{equation}
	R = \zeta \gamma (np-n_i^2) \approx \zeta \gamma np. \label{eqn:Rbr}
\end{equation}
Here, $\gamma$ is the Langevin recombination prefactor, proportional to the charge carrier mobility, and $\zeta$ is the experimentally found reduction factor,\cite{pivrikas2005a,deibel2009,juska2009} which is between $10^{-3}$ and $10^{-1}$. As an upper limit, we used $\zeta=0.1$. The results are shown in Fig.~\ref{fig:j-rescaled}. The minimum of the absolute photocurrent indicates its zero-crossing at $V_s$, where illuminated and dark current equal. Street et al.\cite{street2010} point out that the internal field changes direction at $V_s$. The internal field \emph{in the bulk}, however, changes at the point of optimum symmetry, the quasi-flatband voltage $V_{qfb}$.\cite{limpinsel2010} The current magnitude scales linearly with the generation rate until about 1 sun, despite the recombination being of bimolecular kind. Deviations are only seen for higher illumination intensities. It becomes clear that the voltage-dependent photocurrent does not deviate significantly from the linear scaling with light intensity up to 1 sun---even at voltages close to the quasi-flatband voltage---despite the losses being bimolecular only.

\section{Determination of the $\mu \tau$-Product}\label{sec:mutau}

Street et al.\ present current--voltage characteristics of illuminated polycarbazole/fullerene blend solar cells, and fit them by applying the Hecht expression, Eqn.~(8) of Ref.~\cite{street2010},
\begin{equation}
 	\frac{Q}{Q_0} = \frac{\mu\tau(V_s-V)}{d d'} \left( 1-\exp\left(- \left(\frac{\mu\tau(V_s-V)}{d d'}\right)^{-1}\right)\right).
	\label{eqn:hecht}
\end{equation}
$Q/Q_0$ is the fraction of generated charge that is extracted and not trapped. $d$ is the sample thickness, $d'=d/2$ the depth from which charges have to be collected due to a photogeneration throughout the whole device. $V$ is the applied voltage, $V_s$ the voltage where the photocurrent becomes zero. $F=(V_s-V)/d$, which is an approximation assuming a constant field throughout the device. We define a collection depth $d_c$, from which charges can be extracted before recombining due to their limited lifetime $\tau$,
\begin{equation}
	d_c = \mu \tau F ,
\end{equation}
so that the previous equation becomes
\begin{equation}
 	\frac{Q}{Q_0} = \frac{d_c}{d/2} \left( 1-\exp\left(- \left( \frac{d_c}{d/2} \right)^{-1} \right)\right).
\end{equation}

This equation was fitted to the current--voltage characteristics of the polycarbazole/fullerene solar cells at different light intensities (varied by a factor of 40) by Street et al.\cite{street2010}. The result was presented in Fig.~11 of Ref.~\cite{street2010}, and values for $\mu\tau/(d\cdot(d/2))$ of 6V$^{-1}$ are mentioned in the text, independent of light intensity. The linear scaling of the photocurrent and the light independent value of $\mu\tau/(d\cdot(d/2))$ lead Street et al.\ to the interpretation that the dominant recombination mechanism has to be monomolecular (first order process).

\begin{figure}
	\includegraphics[width=7.0cm]{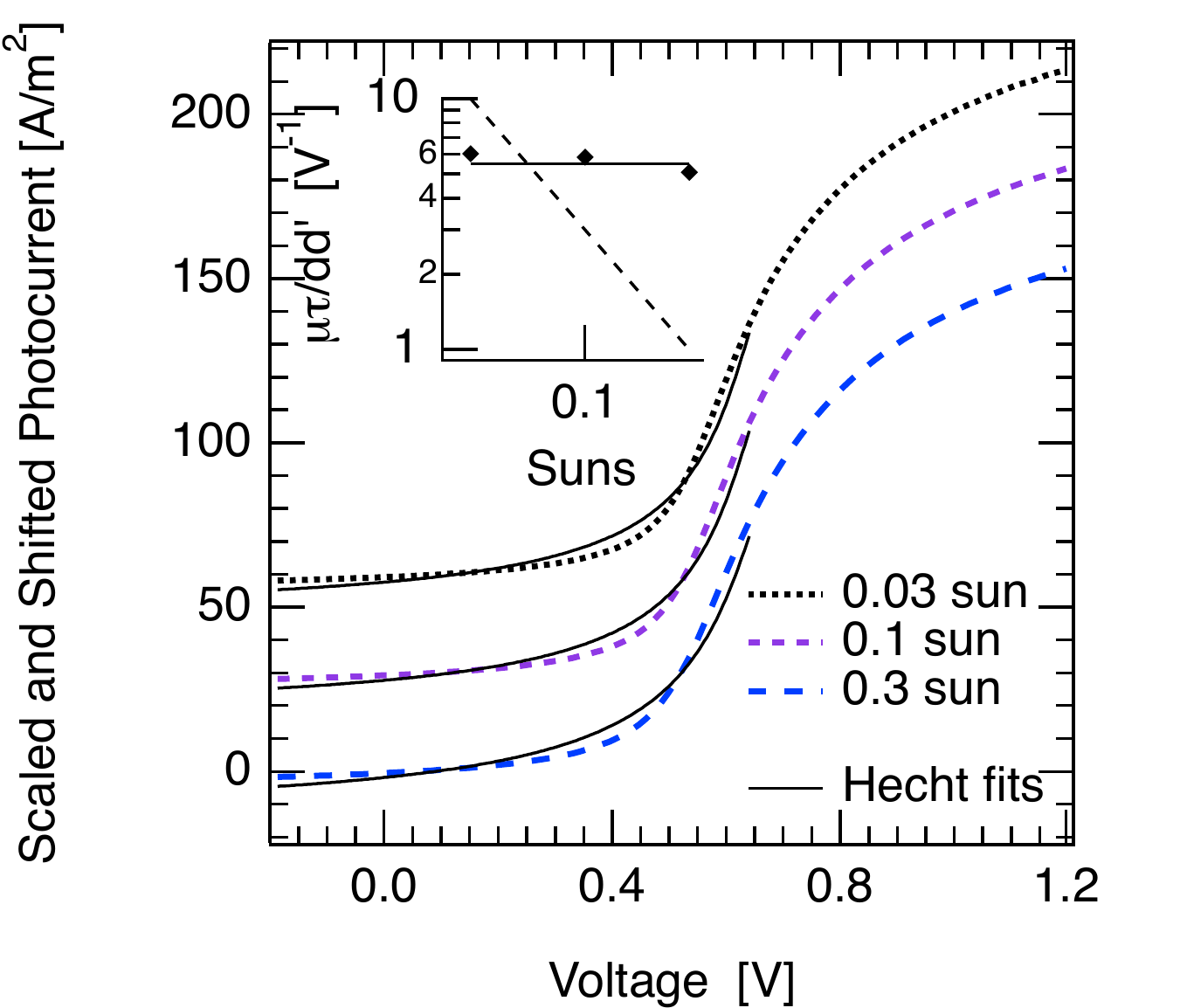}%
	\caption{(Color Online) Hecht Fits of the photocurrent--voltage curves with the three lowest light intensities from Fig.~\ref{fig:j-rescaled} according to Eqn.~(\ref{eqn:hecht}).  We found $\mu\tau/(d\cdot(d/2))\approx5$ to $6$V$^{-1}$ independent of the light intensity.\label{fig:hechtfits}}
\end{figure}

We fitted our simulated photocurrents (Fig.~\ref{fig:j-rescaled}) to Eqn.~(\ref{eqn:hecht}). The fits for 0.3 suns or less are shown in Fig.~\ref{fig:hechtfits}. Street also used illuminations up to about 0.3 suns. Our fits for the photocurrents from 1 to 10 suns became gradually less good with light intensity. The photocurrent for 10 suns does not follow the Hecht equation; similarly, the photocurrent at 1 sun for thicker devices (e.g., 300nm instead of 100nm) cannot be described by Eqn.~(\ref{eqn:hecht}) (not shown). In these cases, the bimolecular losses do become apparent. For the fits shown in Fig.~\ref{fig:hechtfits}, we determined the value $\mu\tau/(d\cdot(d/2))$ to be between 5 and 6V$^{-1}$, very close to the result of 6V$^{-1}$ presented by Street et al.\cite{street2010} in Fig.~11. In the inset of our Fig.~\ref{fig:hechtfits}, the light intensity dependence of $\mu\tau/(d\cdot(d/2))$ is shown. Despite exclusively bimolecular losses being used in the simulation runs, $\mu\tau/(d\cdot(d/2))$ is approximately light intensity independent. 

Previously we published results on temperature dependent measurements of the mobility--lifetime-product by performing time-resolved charge extraction experiments, in which $\mu$ and $\tau$ are extracted simultaneously, but separately. We point out that, although Street et al.\ call their experiments `measured recombination kinetics´, implying time-resolved measurement, they report on quasi-steady-state current voltage measurements using a lock-in amplifier at comparably low frequency. Our experimental data, taken by a time-resolved technique on polythiophene:fullerene solar cells, shows that under short circuit conditions, the photocurrent is not limited by charge recombination.\cite{deibel2009c}: even at temperatures of 200K is $d_c>d$, and strongly increasing with temperature. For annealed devices at 260K under short circuit conditions, $\mu\tau/(d\cdot d/2)$ comes to about 140. We note that these values were taken for a different material system, but highlight the need for performing time-resolved measurements for investigating the mobility--lifetime product.

\section{Discussion and Conclusions}

The interpretation of steady-state current--voltage measurements on polycarbazole/fullerene solar cells lead Street et al.\ to propose first order recombination to be dominant. Therefore, Street et al.\  rule out mechanisms which are of second order, among them bimolecular free carrier recombination---which they call nongeminate exciton recombination, as the free charges first meet to build a  nongeminate charge transfer complex, followed by recombination to the ground state.

We have shown that the determination of the dominant recombination mechanism from steady-state current--voltage measurements alone---calculating the $\mu \tau$-product (Sec.~\ref{sec:mutau}) or considering the light intensity dependence of the photocurrent (Sec.~\ref{sec:jsc})---is ambiguous and almost impossible, at least for illumination levels up to 1 sun. This holds even for voltages close to the quasi flatband case, where recombination should be dominant. The latter behaviour can possibly be explained by the voltage dependent extraction of photogenerated polarons from a bipolar single layer device, as considered by Sokel and Hughes\cite{sokel1982} even without recombination.\cite{mihailetchi2004a,limpinsel2010}

We point out that a wide range of publications have reported on the dominant recombination mechanism in organic solar cells being bimolecular. The studies were based on different techniques in the time or frequency regime.\cite{montanari2002,koster2006,shuttle2008a, shuttle2008b, pivrikas2005a,juska2006,foertig2009,garcia-belmonte2010,mozer2010,deibel2010review} The authors of these studies share the opinion that nongeminate bimolecular recombination is a strongly limiting factor in organic solar cells. In addition, the influence of geminate recombination of singlet excitons and charge transfer excitons can mostly be seen in the saturation current density---for the electric fields seen in working solar cells, these processes are not (singlet excitons) or only weakly (charge transfer excitons) field dependent.\cite{deibel2009a,clarke2010review}

Therefore, we believe that the conclusions of Street et al.\ should be reconsidered, including the assignment of interface recombination as the cause of the first order recombination and the guidelines on how to improve the efficiency of organic solar cells.

%\section*{Acknowledgments}

%\bibliography{Papers}

\end{document}